\documentclass[useAMS,usenatbib]{mn2e}

\usepackage{graphicx}
\usepackage{bm}
\usepackage{natbib}
\usepackage{subfig}
\usepackage{subfloat}
\usepackage{listings}
\usepackage[intlimits]{amsmath}
\usepackage{txfonts}
\usepackage{bm}
\usepackage{amssymb}

\newcommand{\be}{\begin{equation}}
\newcommand{\ee}{\end{equation}}

\newcommand{\ba}{\begin{eqnarray}}
\newcommand{\ea}{\end{eqnarray}}

\title[Hall drift and pulsar braking indices]{Hall drift and the braking indices of young pulsars}

\author[K.N.~Gourgouliatos \& A. Cumming]{\parbox{\textwidth}{K.N.~Gourgouliatos\thanks{E-mail: kostasg@physics.mcgill.ca}$^{1}$, A.~Cumming$^{1}$
}\vspace{0.4cm}\\
\parbox{\textwidth}{ $^{1}$ Department of Physics, McGill University, 3600 rue University, Montr\'eal, Qu\'ebec H3A 2T8, Canada} }

\begin{document}

\date{Accepted -. Received -; in original form -}
\pagerange{\pageref{firstpage}--\pageref{lastpage}} \pubyear{-}
\maketitle

\label{firstpage}

\begin{abstract}
Braking index measurements of young radio pulsars are all smaller than the value expected for spin down by magnetic dipole braking. We investigate magnetic field evolution in the neutron star crust due to Hall drift as an explanation for observed braking indices. Using numerical simulations and a semi-analytic model, we show that a $\approx 10^{14}\ {\rm G}$ quadrupolar toroidal field in the neutron star crust at birth leads to growth of the dipole moment at a rate large enough to agree with measured braking indices. A key factor is the density at which the crust yields to magnetic stresses that build up during the evolution, which sets a characteristic minimum Hall timescale. 
The observed braking indices of pulsars with inferred dipole fields of $\lesssim 10^{13}\ {\rm G}$ can be explained in this picture, although with a significant octupole component needed in some cases. For the stronger field pulsars, those with $B_d\gtrsim 10^{13}\ {\rm G}$, we find that the magnetic stresses in the crust exceed the maximum shear stress before the pulsar reaches its current age, likely quenching the Hall effect. This may have implications for the magnetar activity seen in the high magnetic field radio pulsar PSR~J1846-0258. Observations of braking indices may therefore be a new piece of evidence that neutron stars contain subsurface toroidal fields that are significantly stronger than the dipole field, and may indicate that the Hall effect is important in a wider range of neutron stars than previously thought.
\end{abstract} 

\begin{keywords}
stars: neutron, methods: analytical, methods: numerical, MHD
\end{keywords}

\section{Introduction}
Precise timing of pulsars has shown that they spin down steadily because of magnetic braking \citep{Richards:1969}. The braking index $n=\Omega \ddot{\Omega}/\dot{\Omega}^{2}$, with $\Omega$ the angular frequency and dots denoting time derivatives, is a characteristic quantity of the spin-down evolution of a pulsar. According to the vacuum dipole model by \cite{Deutsch:1955} a pulsar radiates energy at a rate $\dot{E}=-(B_{d}^{2}R_{*}^{6}\Omega^{4}/6 c^{3})\sin^{2}\alpha$, where $B_{d}$ is the dipole field intensity at its magnetic pole, $R_{*}$ the radius, $\alpha$ is the inclination angle between the rotational axis and the magnetic moment and $c$ is the speed of light. Numerical simulations of the magnetosphere find the same dependence on $\Omega$ and $B_{d}$, with somewhat different dependence on $\alpha$ \citep{Spitkovsky:2006}. The loss of kinetic energy is therefore described by
\begin{equation}\label{eq:energy}
{d(I_{m} \Omega^{2})\over dt}=-K B_{d}^{2}\Omega^{4},
\end{equation}
where $I_{m}$ is the moment of inertia onto which the torque acts and $K$ an appropriate numerical factor. Assuming constant $I_{m}$, $K$ and $B_{d}$ gives $\dot{\Omega} \propto \Omega^{3}$, yielding the well-known braking index for magnetic dipole spin-down, $n=3$ \citep{Pacini:1968, GunnOstriker:1969}. 

Phase coherent timing of 8 young pulsars has allowed the measurement of their $\ddot{\Omega}$ (Table 1), giving values of $n<3$ \citep{Manchester:1985, Lyne:1993, Camilo:2000, Livingstone:2007, Espinoza:2011}, smaller than expected in the simple magnetic dipole braking model. A braking index smaller than 3 can arise from either an increase in the torque acting on the pulsar with time (right hand side of eq.~[\ref{eq:energy}]), or a decrease in the moment of inertia with time (left hand side of eq.~[\ref{eq:energy}]). Several interpretations of the observed braking indices have been put forward. One suggestion is that magnetospheric effects change the spin dependence of the torque acting on the pulsar \citep{Contopoulos:2006}, with $n<3$ resulting when the corotating region of the magnetosphere closes within the light cylinder. A fallback disc around a pulsar can provide extra torque \citep{Menou:2001}, however it has been shown recently that their formation is not very likely \citep{Perna:2014}, and only one example, a passive debris disk, has been found in observational searches \citep{Wang:2006}. Alternatively, it has been proposed that the effective moment of inertia of a NS can decrease as normal matter turns into superfluid and decouples from the spin down \citep{Sedrakian:1998, Ho:2012}, or because of glitch activity  \citep{Alpar:2006}. 

A promising avenue is magnetic field evolution \citep{Blandford:1988}. Previous studies considered the reemergence of an assumed buried field by Ohmic diffusion \citep{Muslimov:1996, Ho:2011} or field evolution in the core \citep{Ruderman:1998} to drive a changing dipole moment and deviation from $n=3$. It has also been suggested that magnetic field variations contribute to timing noise in pulsars \citep{Pons:2012, Xie:2013,Tsang:2013}. An evolving magnetic field yields a braking index
\begin{eqnarray}
n= 3- 4\tau_{c} \dot{B}_{d}/B_{d}\,,
\label{EQN}
\end{eqnarray}
where $\tau_{c}=-\Omega/(2\dot{\Omega})$ is the characteristic spin-down age and $\left|B_{d}/\dot{B}_{d}\right|$ is the timescale on which the dipole component of the magnetic field evolves. To set the scale, the measured braking index $n=2.51$ for the Crab pulsar requires that its magnetic dipole moment is increasing at a rate of 1\% per century, $B_d/\dot B_d\approx 10^4\ {\rm years}$. A similar order of magnitude is inferred for other sources (see Figure 1). 
\begin{figure}
\includegraphics[width=\columnwidth]{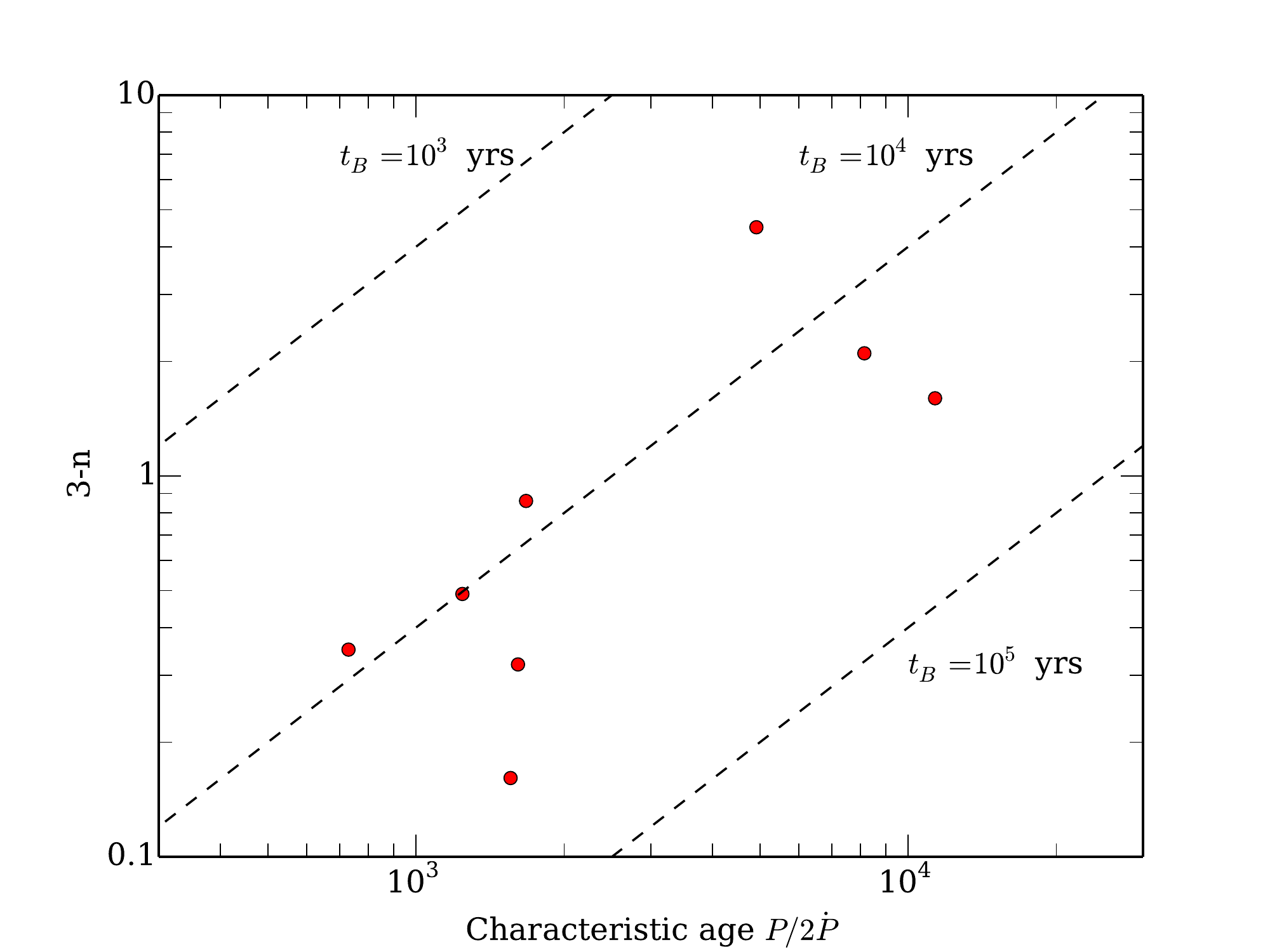}
\caption{The observed deviation from $n=3$ as a function of the characteristic age of the pulsar. The dashed lines show the expected deviation $3-n = 4\tau_c/t_B$ for magnetic field evolution timescales $t_B=10^3$, $10^4$, and $10^5\ {\rm years}$. The increase in the observed values of $3-n$ with $\tau_c$ favours models where $3-n$ depends on the characteristic age. The observations are consistent with a range of $t_B$ of about a factor of 10 around a value $t_B\sim 10^4\ {\rm years}$. The error bars in $3-n$ are not shown, but are $\lesssim 10\%$ (see Table 1) and much smaller than the range of values. 
\label{Figure:dn}}
\end{figure}
\begin{table*}
\resizebox{2.0\columnwidth}{!}{
\label{Table}
\begin{tabular}{ l| r| r| r | r |r | r| r| r  }
Name &   $P$~(s) & $\dot{P}$        & $B_{d}$ & $\tau_{c}$ & n & Age  & $B_{d,i}$ & $B_{\phi ,14}$  \\
	 &            & $(10^{-12}$s/s) &   ($10^{12}$G)  & (yr) & & (yr) & ($10^{12}$G) &   \\        
\hline
J1734-3333  & 	1.17 &  2.28&  $52.3 $ & 8120 & 0.9(2) & $>1,300$	& 45.9 & 13  \\
J1846-0258  & 0.325& 7.08&	48.5 & 729 & 2.65(1) & 1000$^{+3300}_{-100}$ & 41.3 & 32 \\
J1119-6127   & 0.408 &	4.02 & 41.0 & 1,611 &2.684(2)& 7,100$^{+500}_{-2900}$ & 34.5 & 12  \\
B1509-58	     &0.151 & 1.54&	15.4 & 1,556 & 2.839(3)& 	$<$21,000 & 14.2 & 6.8 \\
B0540-69      & 0.0505 & 0.479 & 4.98 &	1,673 & 2.140(9) &	1,000$^{+660}_{-240}$ & 4.16 & 9\\
B0531+21 (Crab) & 0.0331 & 0.423 & 3.79 & 1,242	& 2.51(1) & 960 & 3.30 & 5.6\\
B0833-45 (Vela)   &	0.0893 &	0.125 & 3.38 & 11,300 &1.4(2) &11,000$^{+5000}_{-5600}$ & 1.80 & 12 \\
J0537-6910 & 0.0161 & 0.0518 & 0.93 &4,930 & -1.5(1) & 	2,000$^{+3000}_{-1000}$ & 0.46 & 40 \\
\end{tabular} }
\caption{The parameters of the pulsars with measured braking indices (from \citep{Ho:2012}) and model fits. The last two columns give details of our numerical models that fit each object (section 3). The models have $c_{3}=-c_{1}$, $B_{d,i}$ is the intensity of the dipole component at birth, and $B_{\phi, 14}$ is the maximum intensity of the toroidal component in units of $10^{14}\ {\rm G}$. We could not accommodate the long observed ages of J1119-6127 and B1509-58, in the models presented here, these pulsars reach their observed magnetic fields at ages $\sim$ 3000 and  2000 years from their birth respectively.}
\end{table*}

The neutron star crust is a natural place to look for such short timescale evolution. 
The crust, with thickness of $\sim 1$km, consists of a highly conducting crystal lattice of positively charged nuclei and free electrons. Provided that Lorentz forces can be balanced by the crust elastic forces, the evolution of the magnetic field is dominated by two main processes: Hall drift and Ohmic dissipation \citep{Jones:1988, Goldreich:1992,Pons:2013}. Hall drift is the evolution of the magnetic field as magnetic field lines are advected by the electron fluid and conserves the energy of the magnetic field; while, because of the finite conductivity, some magnetic flux is converted into heat through Ohmic dissipation. The typical timescale for the Hall drift is $t_{Hall}= 4 \pi {\rm e} n_{\rm e} L^{2}/(c B)$, with $n_{\rm e}$ the electron number density, $L$ the scale height for the magnetic field, $B$ the intensity of the magnetic field and ${\rm e}$ the elementary electron charge. This timescale can be short in the outer parts of the crust where $n_{\rm e}$ is small. 

The Hall timescale cannot be made arbitrarily small, however, because at very low electron densities shear stresses in the crust are no longer able to balance the Lorentz forces that develop as the magnetic field evolves. In the very outermost layers, at densities  $n_{\rm e}<n_{\rm e,melt}=2.3\times 10^{31}\ {\rm cm^{-3}}\ T_8^3 (Z/26)^{-5}$, the crust melts and cannot provide shear stress. However, even in the solid layers the magnetic stresses can grow to be large enough that the distortion of the crystal lattice exceeds the breaking strain of the crust.  For a breaking strain of $\epsilon=0.1$ \citep{Horowitz:2009}, setting $B^2/8\pi = \epsilon\mu_s$ where $\mu_s\approx 10^{-2}P$ is the shear modulus of the crust and $P$ the pressure, gives an estimate of the electron density below which the crust will break,
\begin{equation}\label{eq:nb}
n_{\rm e,break}=1.5\times 10^{33}\ {\rm cm^{-3}}\ \left({B\over 10^{13}\ {\rm G}}\right)^{3/2}\left({\epsilon\over 0.1}\right)^{-3/4}.
\end{equation}
The Hall timescale at that density is 
\begin{equation}\label{eq:tHall0}
t_{\rm Hall}\approx 6400\ {\rm years}\ \left({n_{\rm e}\over 10^{33}\ {\rm cm^{-3}}}\right)\left({L^2\over 10^{10}\ {\rm cm^2}}\right)\left({B\over 10^{13}\ {\rm G}}\right)^{-1},
\end{equation}
an interesting match to the $B_d/\dot B_d\approx 10^4\ {\rm years}$ timescale needed to explain the braking index of the Crab pulsar\footnote{We have arbitrarily used a lengthscale of $1\ {\rm km}$, of order the crust thickness in this estimate. We discuss in detail in section 2 the appropriate definition of the Hall timescale. For now, equation (\ref{eq:tHall0}) motivates us to investigate Hall drift in more detail as the source of pulsar braking indices.}.

Recently, we showed that the initial configuration of the magnetic field in the neutron star crust qualitatively affects the subsequent evolution by Hall drift \citep{Gourgouliatos:2014a}. If the magnetic field is in an MHD equilibrium at the time of crust formation (likely since the crust takes many Alfven crossing times to form; \cite{Gourgouliatos:2013}), the dipole field lines are initially advected towards the magnetic pole, increasing the magnetic dipole moment. The opposite behaviour had been seen in previous simulations that chose a different initial condition based on the lowest order Ohmic diffusion eigenmode (e.g.~\cite{Pons:2007}). However, once a quadrupolar toroidal magnetic field of appropriate polarity and intensity is included in the initial state, it temporarily increases the intensity of the dipole component [cf. model B, Fig.~3 \citep{Pons:2007}]

In this paper, we investigate whether growth of the dipole moment driven by the Hall effect could be responsible for the low values of braking index observed in young pulsars. We show that, given a strong subsurface toroidal field component at the time the crust forms, Hall drift can indeed restructure the magnetic field inside the crust of young neutron stars, enhancing the dipole component of the magnetic field, and increasing appropriately the torque to account for the measured braking indices. In section 2, we calculate the growth rate of the dipole moment using numerical simulations and with a semi-analytical model. We  compare to observed braking indices in section 3. We discuss these results and conclude in section 4.

\begin{figure*}
\includegraphics[width=1.8 \columnwidth]{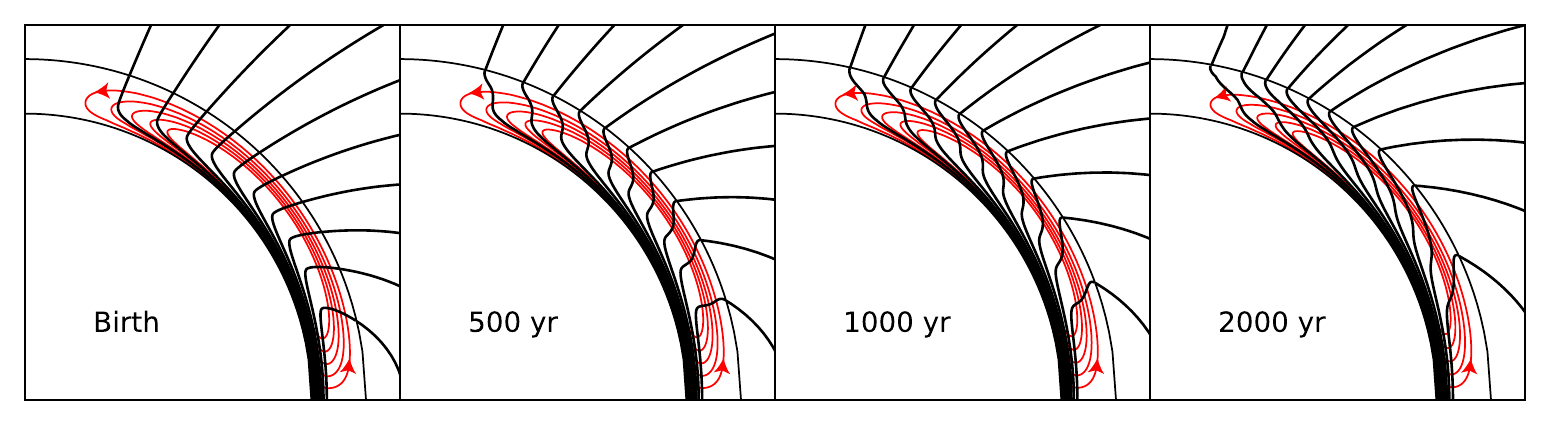}
\caption{The evolution of the poloidal magnetic field under the influence of the toroidal field. The toroidal field is supported by poloidal currents, plotted in red, with the arrows indicating the direction of the motion of the electrons. The moving electrons advect the poloidal field lines, plotted in black, towards the poles, enhancing the dipole moment of the field. The evolution is significant near the surface of the crust, while the field near the base of the crust remains unaffected.}
\label{Figure:1}
\end{figure*}

\section{Growth of the Dipole Moment due to Hall Drift}

In this section, we calculate the time evolution of the dipole moment. We start with numerical simulations of crust field evolution (section 2.1), and then develop a semi-analytic model that captures the main physics and reproduces the behavior observed in our numerical simulations (section 2.2). Of particular importance is the evolution of the toroidal field near the surface of the star, and we discuss this in detail in section 2.3.

\subsection{Numerical simulation of magnetic field evolution in the crust}\label{sec:numerical}

To study the effect of Hall drift on the braking index we simulated the evolution of an axially symmetric field in a NS crust using a finite-difference, forward-time integrating, 2-D axisymmetric scheme presented in \cite{Gourgouliatos:2014a}. 
In the simulations we performed we assumed a NS radius $R_{*}=10^{6}$cm, and a crust thickness of $8\times10^4$cm. We used two electron number density $n_{\rm e}$ profiles for the NS crusts depending on the $B_{d}$ field of the NS we simulated, to approximately take into account the fact that the crust will break at low density as the field evolves. For the pulsars with $B_{d}<10^{13}$G we assumed that $n_{\rm e}=2.8\times10^{40} (1.017-r/R_{*})^{4}$cm$^{-3}$, so that the minimum density at the outermost point of our simulation ($r=R_{*}$) is $2.5\times 10^{33}$cm$^{-3}$ and the $n_{\rm e}$ has a range of three orders of magnitude. For the pulsars with $B_{d}>10^{13}$G, we assumed $n_{\rm e}=1.3\times 10^{40}(1.037 -r/R_{*})^{4}$cm$^{-3}$, so that the minimum density is  $2.5\times 10^{34}$cm$^{-3}$ and has a range of two orders of magnitude. In both models the highest electron density at the base of the crust is $2.5\times 10^{36}$cm$^{-3}$. These profiles are good approximations of more precise crust models by \cite{Cumming:2004} with temperatures $\approx 10^8\ {\rm K}$ and for which the electron density closely follows $n_{\rm e} \propto z^{4}$ where $z$ is the depth from the surface of the crust. 

The choice of initial condition for the magnetic field is of crucial importance. While the late time evolution is towards a particular ``Hall attractor" state \citep{Gourgouliatos:2014b}, the early evolution is a response to the initial structure of the magnetic field. Gourgouliatos \& Cumming (2014b) found that a pure poloidal dipole will, under the action of the Hall effect, first generate a toroidal field in the star; then the poloidal currents associated with the toroidal field advect the poloidal field lines towards the magnetic poles, increasing the dipole moment. We find that to achieve an increase in dipole moment at the rate needed to explain pulsar braking indices, we must short-circuit this process by putting in a strong toroidal field at the beginning of the simulation, when the crust forms. 

A sub-surface toroidal field has previously been suggested to be present in neutron stars. Spin-down measurements of neutron stars determine the dipole moment of the star, but higher order mulitpoles or toroidal field components are not well-constrained. Recent modelling and observations suggest that NSs with relatively weak dipole moments may have substantially larger surface and internal fields \citep{Geppert:2006,Shabaltas:2012,Tiengo:2013, Geppert:2013}. From a theoretical perspective, a newborn NS may have significant multipolar structure in its poloidal field resulting from convection prior to crust formation  \citep{Thompson:2001}, and is likely to host a strong toroidal component as differential rotation following collapse winds poloidal fields \citep{Thompson:1993}. Here we consider a quadrupolar toroidal field as would be generated by radial differential rotation acting on the poloidal dipole.

We considered three different models for the initial poloidal field structure: a dipole $(\ell=1)$, an octupole $(\ell=3)$, and an equal and opposite mixture of a dipole and octupole. We decompose the fields on the surface of the NS in multipoles with coefficients $c_{\ell}=(2\ell +1)/(2\ell +2) \int_{-1}^{1} B_{r} P_{\ell} (\mu) d\mu$, where $P_{\ell}$ the $\ell^{\rm th}$ order Legendre Polynomial, $B_{r}$ the radial component of the magnetic field on the surface of the NS and the cosine of the polar angle $\mu=\cos\theta$. We used a quadrupole $(\ell=2)$ toroidal field, with the same polarity with the dipole. The profile of the toroidal field used was $B_{\phi}(r,\mu)=B_{\phi, 14}~2.5\times10^{18}~(1-r/R_{*})^{2}(r-R_{c})\sin\theta\cos\theta/r~$G, where $R_{c}$ is the crust inner radius, $\theta$ is the angle from the magnetic pole and $B_{\phi,14}$ is the maximum value of the toroidal field in the crust (reached at $\theta=\pi/4$ and $r=9.46\times 10^{5}$cm), in units of $10^{14}$G.

\begin{figure}
\includegraphics[width=\columnwidth]{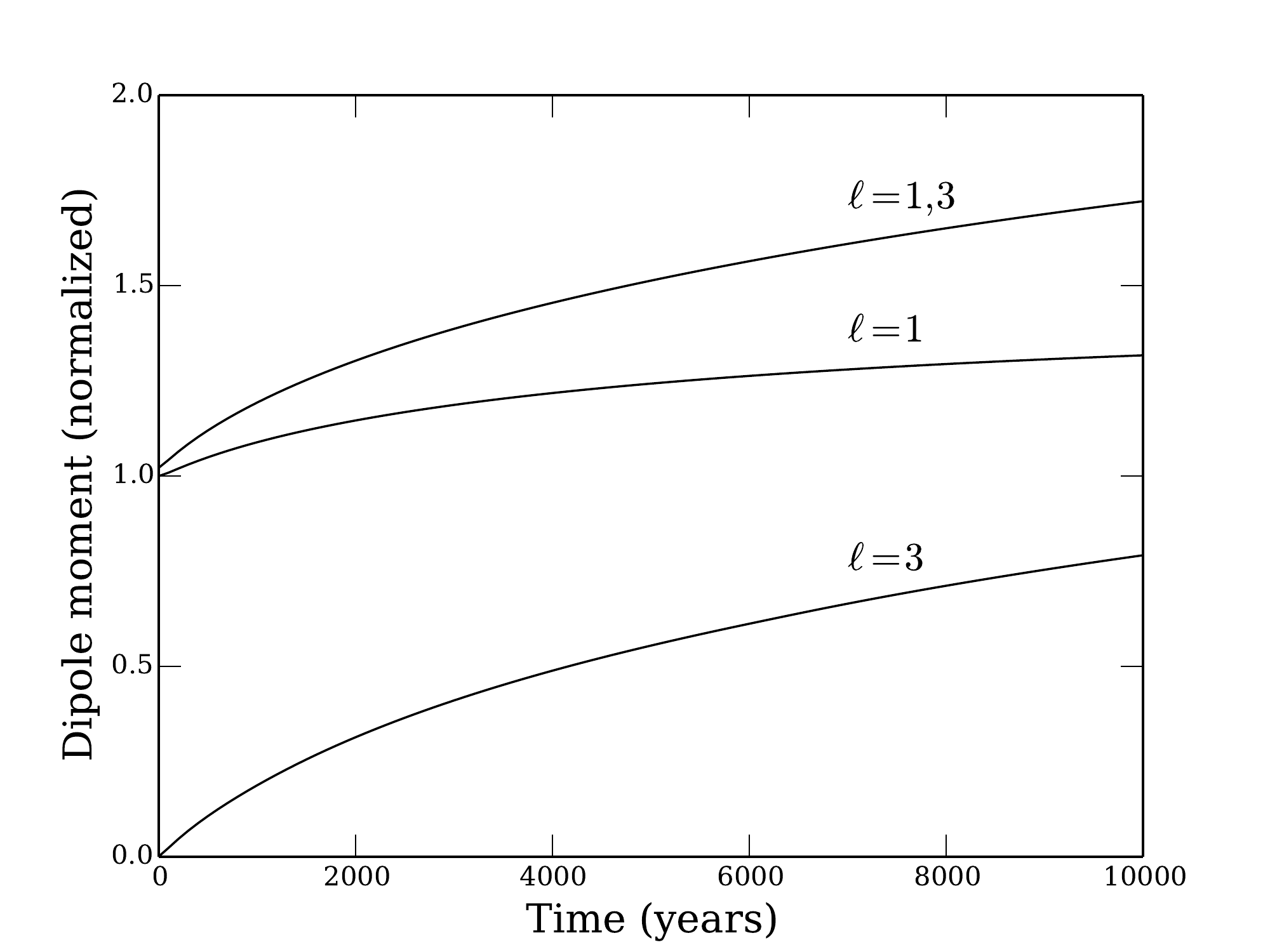}
\caption{The evolution of the dipole moment for three different initial poloidal field configurations, starting with a dipole only, equal and opposite mixture of dipole and octupole, or octupole only. The models shown have $B_{\phi,14}=7$. 
\label{Figure:dipole}}
\end{figure}

We find that a quadrupole toroidal field with the same polarity as the poloidal dipole field winds the poloidal field and stretches the field lines towards the poles (see Figure \ref{Figure:1}). This distortion, caused by advection of the poloidal field lines by the poloidal current loops that support the toroidal field, enhances the dipole moment, and at the same time the field develops an octupole poloidal component of the same polarity. If instead the initial field is of octupole structure, the toroidal field generates a dipole component on the surface of the NS through the same process. Figure \ref{Figure:dipole} shows the evolution of the dipole moment with time for models with $B_{\phi,14}=7$, and the three different initial poloidal field structures. In all three models, the dipole moment changes significantly on timescales of thousands of years. Note that while the redistribution of the poloidal field lines is occurring, Hall drift does not generate new magnetic flux \citep{Reisenegger:2007}. The total magnetic energy does slowly decrease as it is converted to heat through Ohmic dissipation.

\subsection{Semianalytical model of the growth of the dipole moment}

The growth of the dipole moment in the simulations can be understood by considering the passive advection of the poloidal field by the currents associated with the much stronger toroidal field. To do this, we write the axially symmetric magnetic field in terms of scalars $\Psi(r,\mu)$ and $I(r,\mu)$ as $\bm{B}=\nabla \Psi \times \nabla \phi + I \nabla \phi$. The poloidal flux function $\Psi$ evolves according to equation~[7] of \cite{Gourgouliatos:2014a}, 
\begin{equation}
{\partial\Psi\over\partial t} + r^2\sin^2\theta \chi \left(\nabla I\times\nabla\phi\right)\cdot\nabla\Psi = 0,
\end{equation}
where the second term describes the advection of the poloidal field lines by poloidal currents associated with the toroidal magnetic field.
The radial gradient of the toroidal field, $\partial I/\partial r$, dominates, giving an advection equation for $\Psi$,
\begin{equation}\label{eq:advect_psi}
{\partial\Psi\over\partial t}=-v_\mu{\partial\Psi\over\partial \mu}\hspace{1cm} v_\mu = -\left(1-\mu^2\right)\chi{\partial I\over \partial r}.
\end{equation}
Near the surface, where the Hall time is short, $I$ decreases outwards giving $v_\mu>0$ and the poloidal field lines are transported towards the pole. 

Equation (\ref{eq:advect_psi}) allows us to derive evolution equations for the multipole moments. We expand $\Psi$ at the surface of the star as $\Psi = \sum_\ell a_\ell P_\ell^1\left(\mu\right) \sin\theta$, and assuming that the angular dependence of $I$ does not change as it evolves, $\partial I/\partial r = I^\prime(r,t)\mu (1-\mu^2)$, an integral of equation (\ref{eq:advect_psi}) over the surface gives 
\begin{equation}\label{eq:dadt}
{d a_\ell\over dt} = \sum_m {a_mf_{\ell m}\over t_{\rm Hall}(t)}
\end{equation}
where the coefficients 
\begin{equation}
f_{\ell m} = \int^1_{-1}d\mu\ {P_\ell^1\left(\mu\right)\mu\sqrt{1-\mu^2}\over 2}{\partial\over\partial\mu}\left(\sqrt{1-\mu^2}P^1_m\left(\mu\right)\right)
\end{equation}
give the coupling between different $\ell$s, and we define the Hall timescale $t_{\rm Hall}$ by
\begin{equation}\label{eq:tHall}
t_{\rm Hall}^{-1}(t) = 2I^\prime(t) \chi (1-\mu^2)=I^\prime(t) {c\over 2\pi n_{\rm e} e R_{*}^2}.
\end{equation}
The first few coefficients are $f_{11}=1/5$, $f_{13}=-18/35$, $f_{31}=f_{33}=2/15$. 

Equations (\ref{eq:dadt}) are a set of evolution equations for the multipole moments given the time evolution of the background toroidal field. Considering the dipole term only as a first approximation, we find that the dipole moment will grow on a timescale
\begin{equation}\label{eq:tB_dipole}
t_B={a_1\over da_1/dt} = 5 t_{\rm Hall}.
\end{equation}
Equation (\ref{eq:tB_dipole}) matches our simulations that start with a dipole field well; when the field has an octupole component initially, the evolution is faster, closer to $t_B\approx t_{\rm Hall}$. In the dipole case to achieve $t_B\approx 10^4\ {\rm years}$, as needed to match the observed braking index of the Crab for example, requires a Hall time at the surface of $t_{\rm Hall}\approx 2000\ {\rm years}$. 

To solve equations (\ref{eq:dadt}) for the poloidal field evolution, we need the Hall timescale which depends on  $\partial I/\partial r$ (eq.~[\ref{eq:tHall}]). To evaluate the Hall time as a function of time, we have solved the evolution of $I$ with depth in detail by integrating the evolution equation for $I$ including Hall and Ohmic terms 
\begin{equation}\label{eq:dIdt}
{\partial I\over \partial t} = - v_I {\partial I\over \partial r}+{\partial\over \partial r}\left(\eta{\partial I\over\partial r}\right)
\end{equation}
where the advection velocity is $v_I=I\sin\theta\,{\partial \chi/\partial \theta} = -2I\chi\cos\theta=-2I\chi\mu$ (see eq.~[8] of \citealt{Gourgouliatos:2014a}). We follow $I$ in time on a radial grid using the method of lines including the detailed hydrostatic structure of the crust, equation of state, and electrical conductivity as described in \cite{Cumming:2004} (we assume an isothermal crust with $T=10^8\ {\rm K}$). We include the Hall term only for $n_{\rm e}>n_{\rm e,break}$, where $n_{\rm e,break}$ is given by equation~(\ref{eq:nb}). The value of $\partial I/\partial r$ at $n_{\rm e}=n_{\rm e,break}$ was used in equation (\ref{eq:dadt}) to simultaneously solve for the evolution of the multipole moments (we follow odd $\ell$s up to $\ell=9$). We find that this approach reproduces the results from the numerical simulations presented in section \ref{sec:numerical}, including the cases where the initial state has a significant octupole component. This is a useful check on the numerical simulations, as we can include a much larger range of $n_e$ in these 1D calculations of $I$ than in the full 2D numerical simulations.

\begin{figure}
\includegraphics[width=\columnwidth]{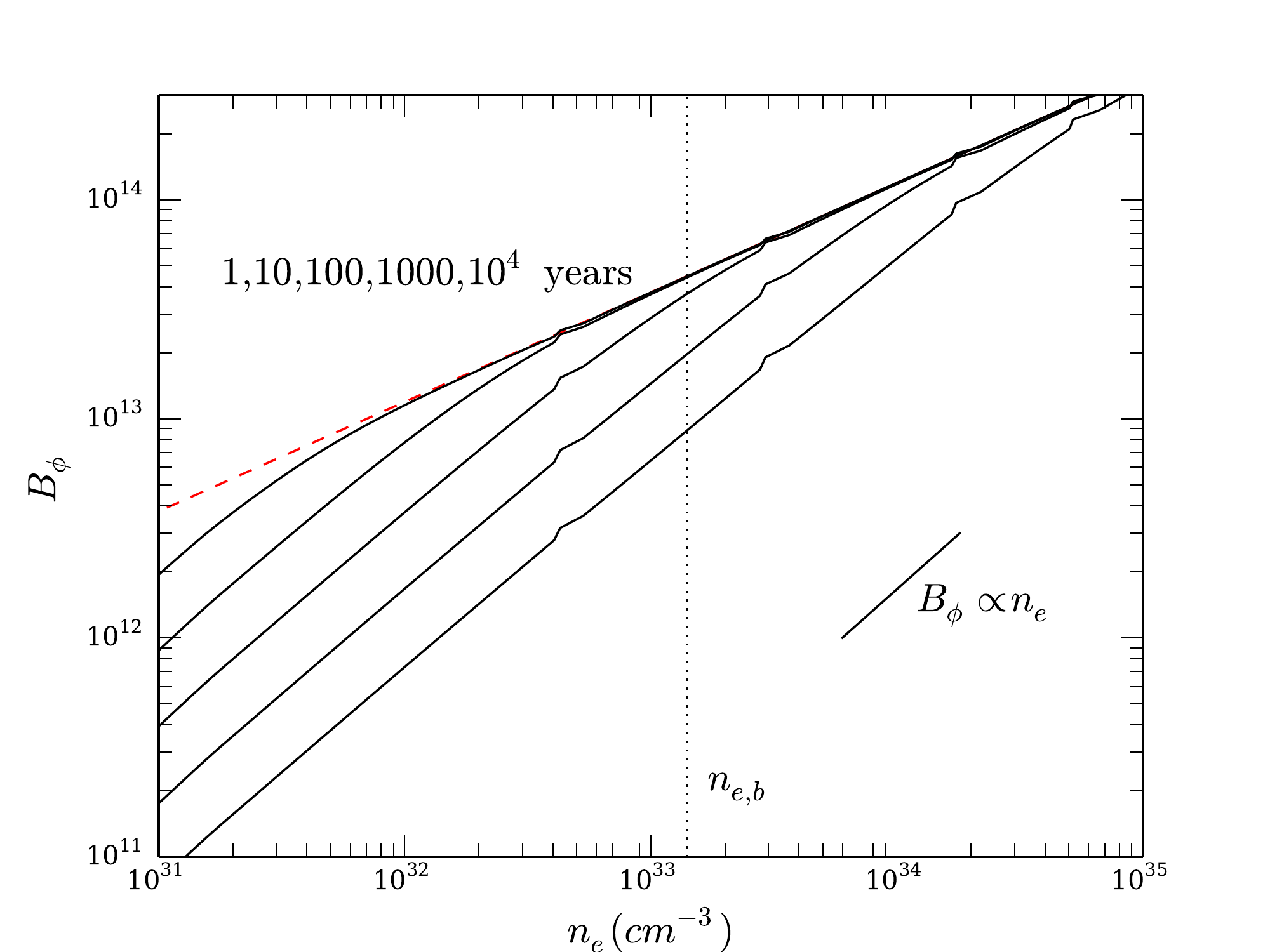}
\caption{The evolution of the radial profile of $B_\phi$ with time for a model that has $B_\phi\propto n_e^{1/2}\propto z^2$ initially, shown as the red dashed line. The profiles are shown at times of 1, 10, 100, 1000, and 10000 years. The kinks in the curves occur at electron capture boundaries where the composition changes as a result of the changing equilibrium nucleus with density. The vertical blue dotted line shows the estimated depth at which crust breaking will occur given the initial field profile (eq.~[\ref{eq:nb}]). Also shown is the scaling $B_\phi\propto n_e$ expected for constant Ohmic flux.\label{Figure:Bprof}}
\end{figure}

\begin{figure}
\includegraphics[width=\columnwidth]{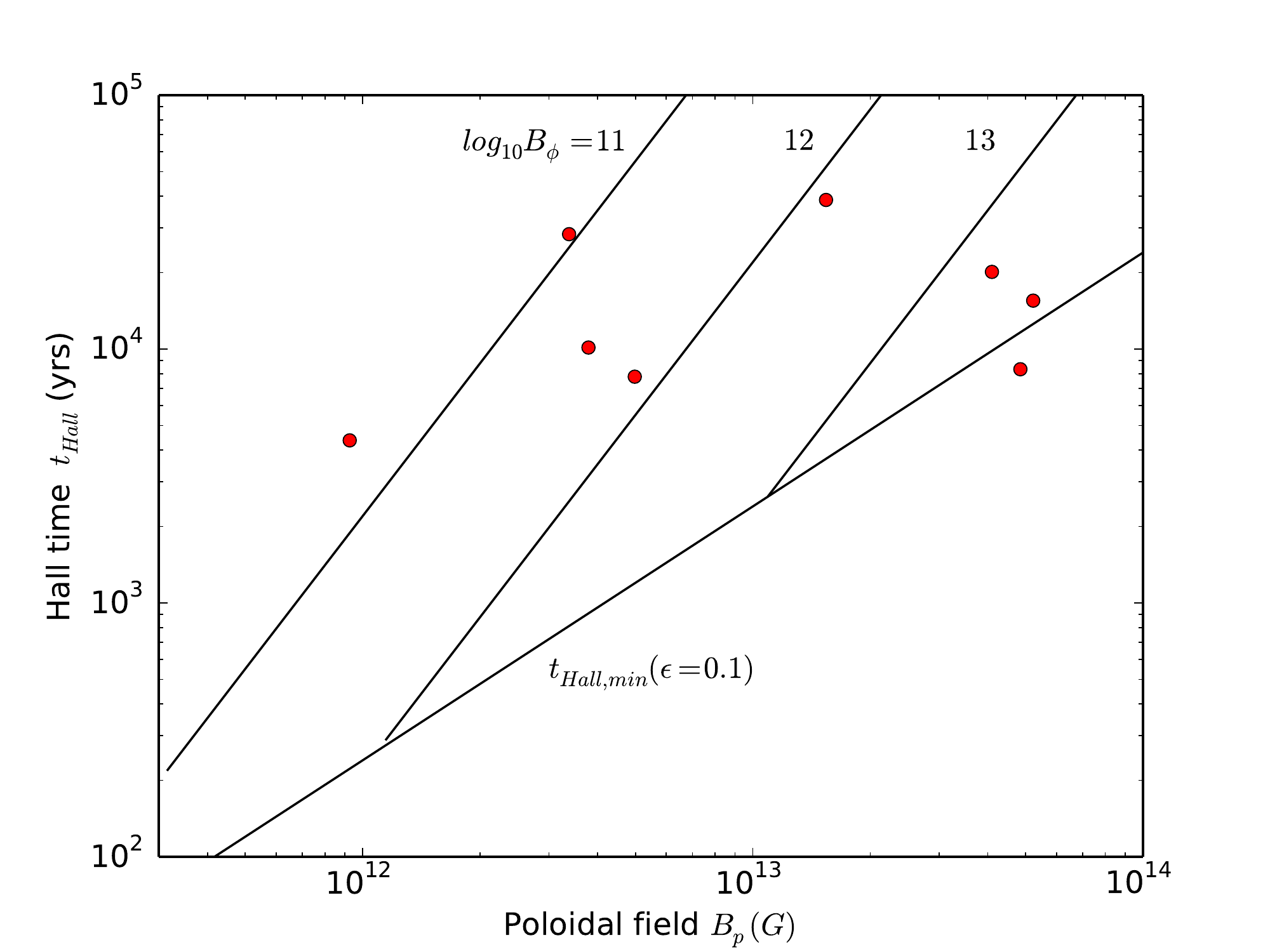}
\caption{Hall timescale as a function of the poloidal and toroidal field, eqs.~(\ref{eq:tHall1}) and (\ref{eq:tHall2}). For a given poloidal field, there is a minimum Hall timescale possible which is when $B_\phi=B_P$ and toroidal stresses begin to dominate. The red circles show the ages and $t_B=4t_c/(3-n)$ for the pulsars with measured braking indices (see Table 1).\label{Figure:tHall}}
\end{figure}

\subsection{Evolution of the radial profile of $B_\phi$ and crust breaking}

Given the crucial role of the toroidal field gradient, $\partial B_\phi/\partial r$, we now discuss its evolution in more detail.  At low densities, the radial gradient of $I$ is set by Ohmic diffusion, which causes the radial profile of the toroidal field to adjust on an Ohmic timescale so that the Ohmic flux $\propto\eta\,dB_\phi/dr$ is constant with depth. In our simulations, we assume an electrical conductivity $\sigma=2.1\times 10^{22}\ n_{{\rm e},33}^{2/3}$ (a close match to the conductivity profile of an isothermal crust at a temperature of $10^8\ {\rm K}$), giving an Ohmic timescale of 
\begin{equation}
t_{\rm Ohm} = {4H^2\over \eta} = 300\ {\rm years}\ n_{{\rm e},33}^{4/3}\left({Y_e\over 0.4}\right)^2,
\end{equation}
where $H=28\ {\rm m}\ n_{{\rm e},33}^{1/3}(Y_e/0.4)$ is the pressure scale height in the outer crust, and $Y_e$ is the electron fraction. At the densities in our simulations $n_{{\rm e},33}=2.5$ and $25$, the Ohmic time is 1000 and 30,000 years respectively. 

In the outer crust, degenerate electrons set the pressure giving $P\propto n_e^{4/3}$. Hydrostatic balance then implies
\begin{equation}
{dP\over dr} = -\rho g\Rightarrow {1\over n_e^{2/3}}{d n_e\over dr} \approx {\rm constant},
\end{equation}
where $g$ is the local gravity.
Constant Ohmic flux therefore implies $dB_\phi/d n_e\approx $ constant, giving $B_\phi\propto n_{\rm e}\propto P^{3/4}$, and
\begin{equation}\label{eq:Iprime}
I^\prime\approx R_* {dB_\phi\over dr}={3\over 4}{RB_\phi\over H}.
\end{equation}
Figure \ref{Figure:Bprof} shows $B_\phi$ as a function of $n_e$ at different times, showing that a region of constant Ohmic flux grows from the surface inwards over time. The numerical solutions closely match the expected $B_\phi\propto n_e$ scaling.

The fact that $\partial I/\partial r$ evolves to a particular value simplifies the calculation of the Hall time. Substituting this value of $I^\prime$ from equation (\ref{eq:Iprime}) into equation (\ref{eq:tHall}) for the Hall time gives the strength of the toroidal field needed to achieve a given Hall timescale,
\begin{equation}\label{eq:Bphi}
B_\phi = 6.3\times 10^{12}\ {\rm G}\ n_{{\rm e},33}^{4/3}\ \left({2000\ {\rm years}\over t_{\rm Hall}}\right)
\end{equation}
where we set $Y_e=0.4$ and $R_*=10\ {\rm km}$. 

The evolution of $B_\phi$ as a function of time depends on its initial radial profile. If $B_\phi$ increases more slowly with density than $B\propto n_{\rm e}$, Ohmic relaxation causes $B_\phi$ to decrease with time at a given $n_{\rm e}$; if $B_\phi$ increases more rapidly with density than $B\propto n_{\rm e}$, Ohmic relaxation will cause $B_\phi$ to increase with time at a given $n_{\rm e}$. Therefore $t_B$ can be either increasing and decreasing with time; the models shown have $t_B$ increasing with time.

We can estimate the shortest possible Hall time by substituting equation (\ref{eq:nb}) for the density at which the crust breaks. In the simple estimate leading to equation (\ref{eq:nb}), we wrote the magnetic stress as $\propto B^2$. In reality, the magnetic stress is not isotropic. When $B_\phi<B_P$ at $n_{\rm e}=n_{\rm e,break}$ ($B_P$ is the poloidal field strength), the appropriate component of the stress to consider is the $B_\theta B_r$ component, which is $\approx B_P^2$. Equations (\ref{eq:Bphi}) and (\ref{eq:nb}) then give the Hall timescale as a function of the poloidal and toroidal field strength
\begin{equation}\label{eq:tHall1}
t_{\rm Hall} = 2200\ {\rm years}\  \left({B_\phi\over 10^{13}\ {\rm G}}\right)^{-1}\left({B_P\over 10^{13}\ {\rm G}}\right)^{2}\left({\epsilon\over 0.1}\right)^{-1}.
\end{equation}
However, if $B_\phi>B_P$, then the appropriate magnetic stress is $\propto B_PB_\phi$, and so we should set $B^2\sim B_\phi B_P$ when estimating the stress that will break the crust. The toroidal field strength then drops out, and we find a minimum Hall timescale 
\begin{equation}\label{eq:tHall2}
t_{{\rm Hall},{\rm min}} = 2200\ {\rm years}\  \left({B_P\over 10^{13}\ {\rm G}}\right)\left({\epsilon\over 0.1}\right)^{-1}.
\end{equation}
The Hall time is shown as a function of $B_P$ in Figure \ref{Figure:tHall}. The pulsars with poloidal fields $B_d<10^{13}\ {\rm G}$ have $t_B$ much longer than $t_{\rm Hall,min}$ and are easily accommodated without breaking the crust. However, the pulsars with $B_d>10^{13}\ {\rm G}$ have $t_B$ close to $t_{\rm Hall,min}$ and our simulations exceed the breaking strain at low densities as we describe in the next section.

\section{Comparison with Observations}

\begin{figure}
\includegraphics[width=\columnwidth]{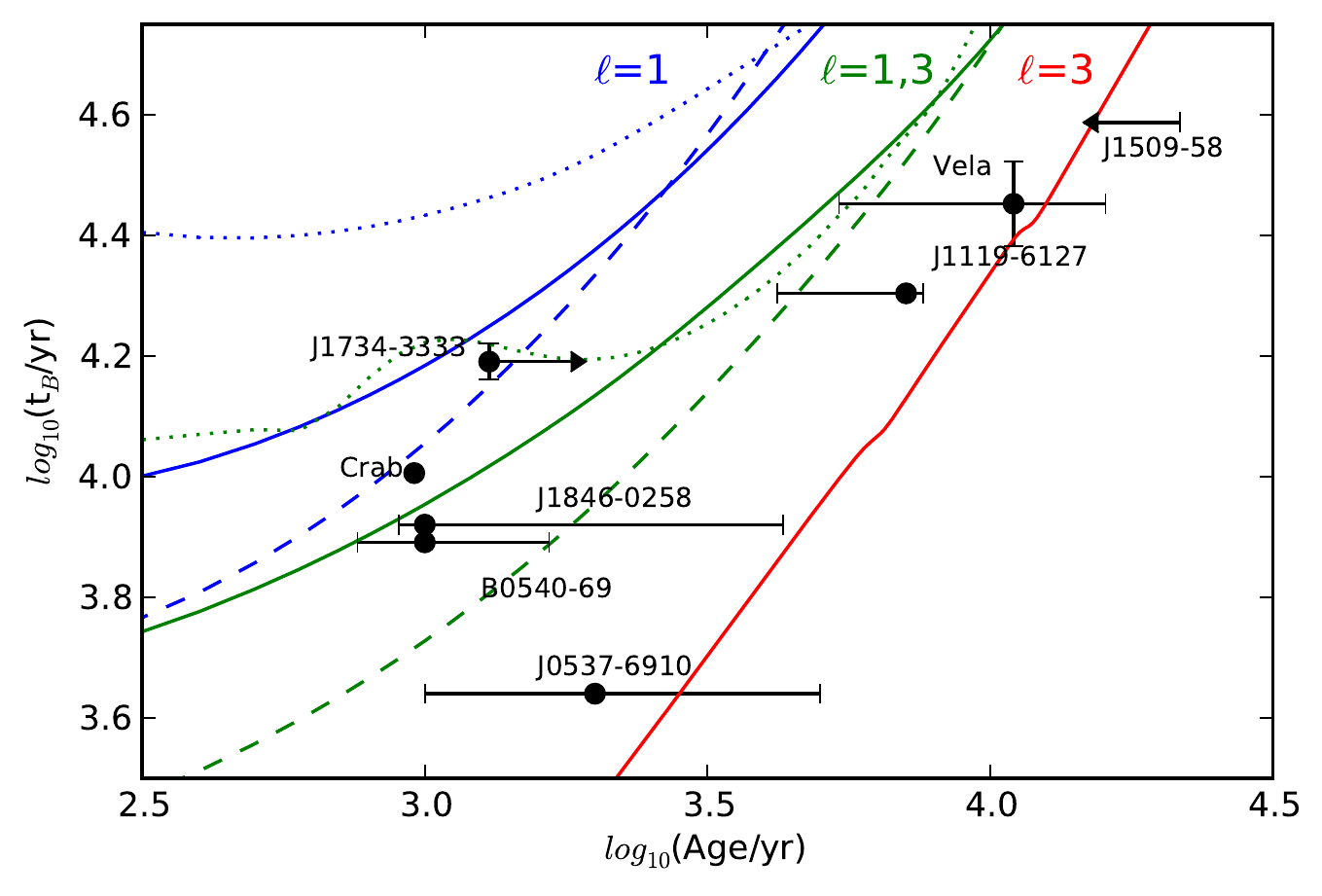}
\caption{The observed $B/\dot{B}$ with pulsar age, for a variety of combinations of poloidal and toroidal fields, and the values  of $(3-n)/4\tau_{c}$ for the eight pulsars whose braking index is known through phase coherent timing, shown in Table 1. The blue lines correspond to a dipole ($\ell=1$) initial field, the green to a mixed dipole and octupole ($\ell=3$), and the red to an octupole only initial field. The solid lines correspond to models with minimum electron density $n_{\rm e}$ of $2.5\times 10^{33}$cm$^{-3}$, and $B_{\phi,14}=7$, while the dashed lines have $B_{\phi,14}=14$. The dotted lines correspond to a crust with minimum $n_{\rm e}$ of $2.5\times10^{34}$cm$^{-3}$ and $B_{\phi,14}=28$. We have used the latter crust models in pulsars with dipole fields above $10^{13}$G, to account the fact that stronger magnetic fields cannot take advantage of all the density range of the crust, as they may deform the outer layers because of strong Lorentz forces.}
\label{Figure:2}
\end{figure}

Having shown that Hall drift associated with a strong toroidal field in the crust leads to growth of the dipole moment on interesting timescales, we now compare the simulations with observed braking indices. The magnetic fields of the pulsars with measured braking indices are in the range $10^{12}$--$5\times 10^{13}$G, and their estimated ages are $\approx 10^{3}$--$10^{4}$ years (see Table 1). To remove the dependence on $\tau_{c}$ (see Figure \ref{Figure:dn}), we solve equation~(\ref{EQN}) to obtain a measured $B_{d}/\dot{B}_{d}$ for each pulsar, and plot it as a function of the pulsar's age in Figure \ref{Figure:2}. A similar diagram was shown by \cite{Ho:2012}, but for moment of inertia evolution rather than magnetic field evolution.

The curves in Figure \ref{Figure:2} show the calculated evolution timescale for different choices of toroidal field strength and poloidal field geometry and strength. The evolution time depends mainly on the toroidal magnetic field (see eq.~[\ref{eq:tHall1}]) and the multipole decomposition of the poloidal field (eq.~[\ref{eq:tB_dipole}] and related discussion). In particular, a stronger toroidal field leads to faster evolution, and the presence of higher order poloidal multipoles provides a reservoir of magnetic flux that can be transferred to the dipole component. 

Figure \ref{Figure:2} shows that a toroidal field strength in the range $10^{13}$--$10^{14}$G in the outer $\approx 100\ {\rm m}$ of the crust gives a good match to the observed braking indices. An initial dipole field is able to reproduce the observations of the Crab, PSR~J1734-3333, PSR~J1846-0258, and PSR~B0549-69. The remaining 4 pulsars require an octupole component to be present to boost the dipole growth rate.

We also find that crust breaking occurs in models of the pulsars with higher poloidal fields $\gtrsim 10^{13}\ {\rm G}$. As we described in section 2, our simulations take into account crust breaking in an approximate way by considering two crust models with differing minimum densities. But in addition, we calculate the magnetic stresses that develop during each simulation following \cite{Perna:2011}, and compare them to the maximum stress that can be accommodated by the solid, to check whether the crust exceeds the breaking strain. For pulsars with stronger fields $\gtrsim 10^{13}\ {\rm G}$, models which match the observed braking indices at the right ages exceed the breaking strain of the crust at low densities by factors of several. 

\begin{figure}
\centering
\includegraphics[width=\columnwidth]{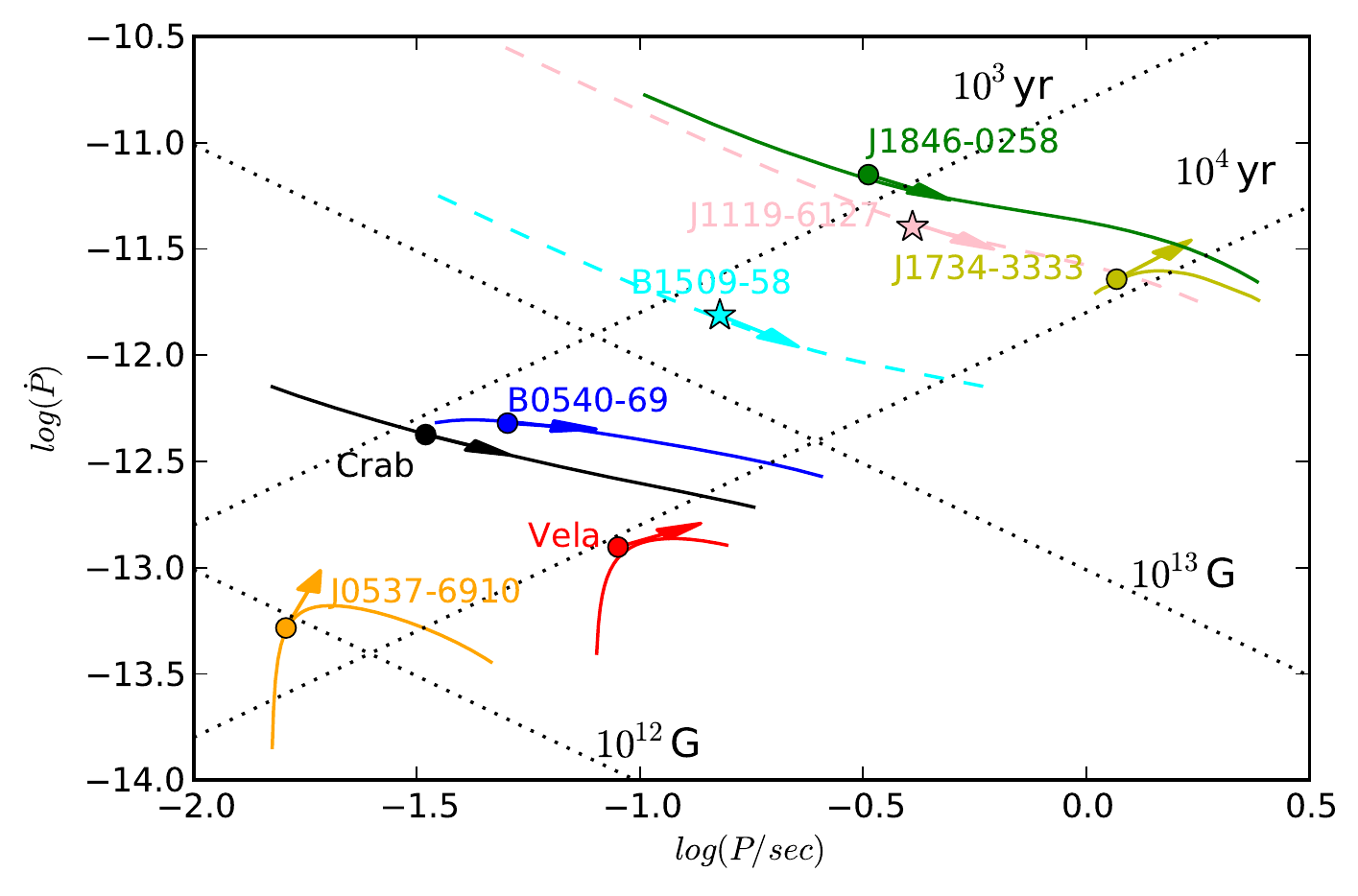}
\caption{$P$-$\dot{P}$ diagram for the eight pulsars for which the braking index is measured, taking into account the magnetic field evolution. The points correspond to the observed $P$ and $\dot{P}$. The direction of the arrow is related to the braking index \citep{Espinoza:2011} $\tan\omega=2-n$, where $\omega$ is the angle with horizontal axis. Examples of evolutionary tracks are plotted, which evolve to the right values of magnetic field and braking index at the pulsar's current age, assuming an initial magnetic field in the ranges shown in Figure \ref{Figure:2}. $P_{0}$ is solved for, given the evolution of the magnetic field and the pulsars position in the $P$-$\dot{P}$ diagram. The tracks of B1509-58 and J1119-6127 are shown as dashed lines, as their estimated ages are much longer than the ones that can be accommodated by a steady spin-down over the pulsar's lifetime.}
\label{Figure:3}
\end{figure}

In Figure \ref{Figure:3} we show models of the evolution in the $P$--$\dot P$ diagram. For each pulsar, we choose the initial poloidal and toroidal magnetic fields (see Table 1) so that the subsequent Hall evolution leads to the observed magnetic field and braking index at the age of the pulsar. Two pulsars, B1509-58 and J1119-6127 (dashed lines in Fig.~\ref{Figure:3}) have estimated ages that are much longer than the ones that can be accommodated by a steady spin-down over the pulsar's lifetime. In those two cases, we show models in which the current age of the pulsar is $\approx$ 2000 and 3000 years respectively, in which case we are able to match the observed values of $P$, $\dot P$, and $n$.


\section{Discussion and Conclusions}

We have shown that an internal quadrupolar toroidal magnetic field of strength $\approx 10^{14}\ {\rm G}$ in young neutron stars leads to an increase of the dipole moment, driven by the Hall effect. The relevant timescale is the Hall timescale at the lowest density in the crust at which the Hall effect can operate, most likely set by yielding of the solid under the magnetic stresses that develop as the field evolves. This depth sets the minimum timescale for field evolution, and is quickly populated with toroidal field by ohmic diffusion,  so that the growth of the dipole moment occurs on a characteristic timescale associated with this depth (eq.~[\ref{eq:tHall1}]).

The observed braking indices of pulsars with inferred dipole fields of $\lesssim 10^{13}\ {\rm G}$ can be accommodated in these models, although with a significant octupole component needed in some cases. For the stronger field pulsars, those with $B_d\gtrsim 10^{13}\ {\rm G}$, we find that the magnetic stresses in the crust exceed the maximum shear stress before the pulsar reaches its current age. Therefore, it is not clear whether Hall drift can explain the braking indices of the higher field pulsars: in the limit where the crust cannot support shear stress, ie. behaves as a liquid, the Hall effect would be expected to effectively switch off, with hydrodynamic motions shorting out the Hall electric fields. It is worth noting that we have assumed a breaking strain $\epsilon\sim 0.1$ \citep{Horowitz:2009}. If the breaking strain is significantly smaller than this, crust yielding would play a role even in the weaker poloidal field pulsars. Further modelling of crust yielding and its back-reaction on the magnetic field evolution is needed to accurately follow the evolution in higher field pulsars. The fact that the crust exceeds its breaking strain due to Hall evolution could explain the magnetar activity observed in PSR~J1846-0258 and other high magnetic field radio pulsars \citep{Gavriil:2008}. 

This calculation focuses in the early evolution of the magnetic field structure of neutron stars ($\sim10^{4}$yr), during which thermal feedback does not have an important effect, since the relevant timescale is much longer \citep{Aguilera:2008, Vigano:2013}. Following the early evolution the dipole component magnetic field has an overall trend to decrease as the crustal magnetic field is Ohmically dissipated, which is consistent with the suggested magnetic field decay from population synthesis studies \citep{Igoshev:2014, Gullon:2014}. While the magnetic field decays, the dipole component of the magnetic field may temporarily increase, because of whistler wave oscillations \citep{Gourgouliatos:2014a}, which has been used to explain the second period derivative measurements inferred by timing noise \citep{Zhang:2012}. 

Magnetic field studies in neutron stars, including the one presented in this work, are constrained to axially symmetric models. The extension to non-axially symmetric calculations may open different paths in the magnetic field evolution and shed light to the role of instabilities and turbulent cascade. These issues stress the importance for the development of a 3-D code that will address such questions.

Even if Hall evolution is not enough to explain the entire deviation of the braking index from $n=3$, a mild toroidal field can still have a significant impact on the spin evolution, and should be taken into account in other interpretations of braking indices. For example, \cite{Ho:2012} suggested that braking index measurements could be used to determine the rate at which the mass of superfluid in the neutron star core is increasing as the star cools. The fact that magnetic field evolution may be changing the dipole moment, even at a low level, will impact constraints on the neutron superfluid component, and therefore NS equation of state, derived in this way.

A larger sample of observed braking indices is needed to test models. Unfortunately, the sample is not likely to increase substantially in size in the near future. An alternative way to test models is a measurement of the second braking index $p=(\Omega^{2}/\dot{\Omega}^{3})\dddot{\Omega}$ \citep{Blandford:1988}. In the context of magnetic field evolution models, the sign of $p$ depends on whether the magnetic field growth is accelerating or decelerating. \cite{Livingstone:2005} were able to measure $p$ for PSR B1509-58, finding $p=18.3\pm2.9$, the central value of which implies that $\ddot{B}/{B}>0$, but is also consistent with $\ddot{B}/B<0$ within $2\sigma$.  Similar conclusions can be inferred by the measurement of the second braking index of the Crab pulsar \citep{Lyne:1988, Lyne:1993}. In the models presented here, measurements of $p$ constrain the initial profile of the toroidal field, specifically the dependence of $B_\phi$ on electron density. 

We have made a particular choice for the toroidal field geometry, a quadrupole, in order to achieve growth of the magnetic dipole moment. A quadrupolar toroidal field arises naturally in Hall evolution of a poloidal dipole field; however, in order to grow the dipole quickly enough, we introduce the toroidal field at the beginning of the simulation. There have been several suggestions that young neutron stars host internal toroidal fields (e.g.~\cite{Geppert:2006, Thompson:1993,Shabaltas:2012,Tiengo:2013}), and the particular choice we made here would arise from radial differential rotation acting on a poloidal dipole field before the crust solidifies, but the unknown state of the magnetic field at the time of crust formation remains a major uncertainty.

The idea that Hall drift may be operating, and have observable consequences, in pulsars is intriguing. 
 The Hall effect is usually discussed as the mechanism underlying observed activity in magnetars \citep{Perna:2011} but the observed braking indices of young pulsars may be telling us that the Hall effect plays a role in a much wider range of systems, including neutron stars with significantly smaller dipole fields than magnetars.

\section*{Acknowledgements}

We thank Dave Tsang, Vicky Kaspi, Hongjun An and Rob Ferdman for insightful discussions. KNG was supported by the Centre de Recherche en Astrophysique du Qu\'ebec. AC is supported by an NSERC Discovery Grant and is an Associate Member of the CIFAR Cosmology and Gravity program.

\bibliographystyle{mnras}
\bibliography{BibTex.bib}

\end{document}